# On the Combined Use of Extrinsic Semantic Resources for Medical Information Search

Mohammed Maree[1, *], Israa Noor[2], Khaled Rabayah[1], Mohammed Belkhatir[3], and Saadat M. Alhashmi[4]

[1]Department of Information Technology, Faculty of Engineering and Information Technology, Arab American University, P.O Box 240 Jenin, 13 Zababdeh, Palestine
[2]Department of Computer Science, Faculty of Engineering and Information Technology, Arab American University, P.O Box 240 Jenin, 13 Zababdeh, Palestine
[3]Faculty of Computer Science, University of Lyon, Lyon, France
[4]Department of Management Information Systems, College of Business Administration, University of Sharjah, Sharjah, UAE

Corresponding author: Mohammed Maree (e-mail: mohammed.maree@aaup.edu).

**ABSTRACT** Semantic concepts and relations encoded in domain-specific ontologies and other medical semantic resources play a crucial role in deciphering terms in medical queries and documents. The exploitation of these resources for tackling the semantic gap issue has been widely studied in the literature. However, there are challenges that hinder their widespread use in real-world applications. Among these challenges is the insufficient knowledge individually encoded in existing medical ontologies, which is magnified when users express their information needs using long-winded natural language queries. In this context, many of the users' query terms are either unrecognized by the used ontologies, or cause retrieving false positives that degrade the quality of current medical information search approaches. In this article, we explore the combination of multiple extrinsic semantic resources in the development of a full-fledged medical information search framework to: i) highlight and expand head medical concepts in verbose medical queries (i.e. concepts among query terms that significantly contribute to the informativeness and intent of a given query), ii) build semantically-enhanced inverted index documents, iii) contribute to a heuristical weighting technique in the query-document matching process. To demonstrate the effectiveness of the proposed approach, we conducted several experiments over the CLEF e-Health 2014 dataset. Findings indicate that the proposed method combining several extrinsic semantic resources proved to be more effective than related approaches in terms of precision measure.

## I. INTRODUCTION

Contrarily to generic web-based search queries that tend to be short, medical queries are long-winded with a reported average length of five terms when examining the query log of an Electronic Health Record search engine [1]. Furthermore, their processing through statistical techniques alone appears insufficient since they encompass several domain-specific medical concepts [2] that require making use of extrinsic knowledge for their deciphering [3]. This forms a crucial challenge for Medical Information Retrieval (MIR) systems that aim to find matches between medical documents and their corresponding queries in the same domain [4-6] and motivates the use for language resources when further expanding these queries [7, 8]. Recently, MIR systems have shifted to exploiting medical semantic resources and ontologies in an attempt to capture knowledge in this domain through formally and explicitly defining medical concepts, instances, as well as semantic and taxonomic relations that link related concepts. Several examples of these resources can be found at the BioPortal[1] website. However, despite the constant growth of current medical semantic resources, they are still insufficient in terms of their domain coverage at both breadth and depth levels (i.e. they formally encode domain conceptualizations at different granularity levels) [9-11]. The main reason behind this limitation is referred to the fact that they are being developed by experts who adopt different standards and use various languages to describe them [12]. Indeed, the incompleteness of the captured semantic information can substantially affect the quality of systems relying on them [13]. On the other hand, MIR systems face another important challenge that has a major impact on their effectiveness. This challenge is manifested by the diversity of users, their information needs and their background knowledge in the medical domain [5, 14, 15]. Addressing each of these challenges plays a crucial role in the way medical query processing and expansion techniques are developed, and has a direct impact on the quality of the retrieved results by MIR systems [16-19]. Starting from this position, we propose a semantics-based MIR system that aims to improve the quality of the returned results through incorporating multiple medical semantic resources and query expansion techniques. In particular, we use the UMLS Metathesaurus [20], which is a large-scale biomedical thesaurus that provides explicit specifications of biomedical knowledge, consisting of concepts classified by semantic type, in addition to the hypernymy-hyponymy relation and other non-hierarchical relationships among the concepts. We use two resources to exploit the UMLS Metathesaurus in our proposed system:

1. The MetaMap tool which maps biomedical texts to the UMLS Metathesaurus. It locates all UMLS concepts associated with terms in biomedical texts using knowledge intensive methods based on symbolic, natural language processing and computational linguistic techniques in the same manner as proposed in [21].
2. The MRDEF relational table that contains UMLS concept definitions from multiple medical semantic resources. In our approach, we use the 'MSH' source which is obtained from the Medical Subject Headings (MeSH) thesaurus and contains 29,244 different concepts.

---

[1] https://bioportal.bioontology.org/

We furthermore consider the UMLS SPECIALIST lexicon (a.k.a. UMLS lexicon) that is provided by the National Library of Medicine (NLM) [22]. This lexicon is one of the richest available sources of medical lexical information and has been employed for the purpose of analyzing medical text [23]. In the context of our work, the UMLS lexicon is used to carry out the following tasks:

1. Extract medical acronyms and abbreviations from user queries.
2. Expand the extracted acronyms and abbreviations into their full representations and use them to reformulate the original queries.
3. Expand medical terms in the original queries by finding their related medical synonyms using the exploited medical semantic resource.

In this context, when a user submits a medical query, the proposed system analyzes the query to identify head concepts in addition to other supportive query terms and enriches them with semantically-relevant terms derived from the collective integration of results from the used resources. The expanded queries are then matched with their corresponding medical documents. In our approach, for each medical document, a semantics-based inverted index is automatically constructed through utilizing the same medical resources that we employ for enriching user queries. By carrying out this step, the matching task is performed at the semantics-level wherein medical documents are ranked according to their semantic closeness to their relevant queries. The main contributions of our proposition are summarized as follows:

1. Employing multiple medical semantic resources for:
    a. Identifying and enriching head concepts of medical queries with semantically and taxonomically related terms. In this context and unlike conventional methods that attempt to enrich queries using individual semantic resources, we employ multiple resources that collectively suggest enrichment candidates.
    b. Constructing semantically-enriched inverted files that encode the latent semantic information within the content of medical documents. Accordingly, rather than relying on representative keywords, our method constructs indexes comprising additional semantic dimensions that are utilized for retrieval purposes.
2. Identifying and re-weighting medical query terms based on the employed semantic resources. In this context, medical terms are assigned higher weights against other supportive terms. We demonstrate the importance of this step and its impact on the quality of the proposed system in the experimental evaluation section.

In Section 2, we review related literature. In Section 3, we describe the overall organization of the proposed framework and detail the query and document processing steps as well as the matching and weighting techniques. We present the empirical setup and the produced results of the conducted experiments in Section 4. In Section 5, we discuss the conclusions and outline the future extensions to our current work.

## II. Related Work

The utilization of Natural Language Processing (NLP) techniques and medical semantic resources for processing medical queries has been at the heart of MIR systems for years [2, 5, 6, 14-19, 24-35]. For instance, Zhu and Carterette proposed a medical record search system for identifying cohorts required in clinical studies [33]. To do so, the authors employed a query-adaptive weighting method that can dynamically aggregate and score evidence within multiple medical reports. They proposed using a number of features such as length of the query, number of concepts in the query, broad/narrow query concepts, etc.

that can be exploited to assign weights for medical concepts in the supplied queries. Medical concepts are detected using MetaMap [21], a medical NLP tool developed by the National Library of Medicine (NLM) to map biomedical text to concepts in the Unified Medical Language System (UMLS) Metathesaurus. The authors cross-validated that their weighting method is better than a fixed-weighting method across several evaluation metrics. Though, according to the authors, the improvement was not statistically significant, and the proposed method had the potential to be further improved by incorporating other useful features or by using advanced prediction models. In light of this argument, we would like to also point out that using the UMLS Metathesaurus alone for mapping medical concepts is not sufficient due to the limited domain coverage of this knowledge base. In a similar line of research, Martinez et al. proposed to automatically expand medical queries based on the concepts and relations included in the UMLS [34]. The query expansion method relied on an algorithm known as *Personalized PageRank*, which runs over the graph representation of the UMLS structure. The intention of using this algorithm was to initialize the probability distribution of the UMLS graph with the terms highlighted in the query to identify relevant terms, which can be used to expand the query for improving the retrieval of relevant medical documents. To demonstrate the effectiveness of their proposed approach, the authors conducted experiments using the TREC Medical Record track, showing improvements in both the 2011 and 2012 datasets over baseline methods. However, we argue that despite the achieved improvement by the proposed approach, the reliance only on the UMLS knowledge base is not sufficient. This is mainly because of the domain knowledge incompleteness problem [9, 15], which is also acknowledged by the authors of the proposed approach.

In a recent work detailed in [6], the authors proposed an automatic medical query expansion method that starts by identifying key terms (i.e. the most effective candidate expansion terms among the query terms) to be used in the matching and retrieval process. To identify key terms, the authors re-used the method proposed in their previous work [2]. Using this method, they located all the contexts in the original document collections that matched the contexts of the key terms in verbose queries. Although the proposed method proved to be efficient in accomplishing the matching task, its effectiveness was hindered by the following facts. First, as stated by the authors, query terms can be single terms or phrases. The authors referred to these types of terms as *key terms* without considering their semantic dimensions that could have an impact on the overall quality of the proposed method. For instance, the authors did not consider the semantic relations that may exist between key terms. Also, they ignored the synonyms as well as other lexically-related terms to each key term. Second, the process of locating all contexts in the document collections may lead to retrieving several false positives as the method relied on keyword overlap between the extracted contexts.

Stanton and his colleagues explored the scenarios wherein a user expresses his information needs using many words to describe a certain symptom [14]. To do this, they proposed a supervised machine learning approach to link terms among the given queries to their corresponding medical concepts. In the context of their work, they first obtained the formal definitions of diseases using medical semantic resources in an attempt to reformulate queries through incorporating the derived medical concepts and their definitions. Although the proposed approach achieved an improvement in mapping symptoms to the proper relevant disease/s, the authors ignored other query term types (those that do not belong to symptoms and diseases) such as laboratory tests, medical devices, etc. In a similar line of research, Shen et al. proposed the *bag-of-concepts* model to identify medical concepts in user queries through exploiting medical knowledge resources [24]. To retrieve medical documents, they used the selected concepts and their mapping entities in the used resources. However, the proposed approach was hindered by two obstacles. First, all non-medical query terms were ignored in the proposed retrieval process. Second, due to limited domain coverage issues, many concepts highlighted in the user queries were not recognized by the used medical knowledge resources. In a similar work by Choi and Choi, the authors proposed a concept-based query expansion model using selective query concepts [25]. In this context, *discharge summary reports* (defined as the resources queries have been built from) from the CLEF eHealth14 dataset and UMLS were used to extract and expand medical concepts. Other concepts that were not in the discharge summary reports were ignored. The proposed system demonstrated minimal improvement on the quality of the produced results because of two reasons. First, the authors did not consider compound terms and stopwords that may exist in the queries and medical documents. Second, they restricted the expansion scope to query-related discharge summary reports provided in the dataset. However, such reports were provided as example results only. On the other hand, Goeuriot et al. focused on using local resources for query reformulation rather than using external medical semantic resources and NLP techniques [35]. In this context, the authors used the Pseudo Relevance Feedback (PRF) model for query reformulation. To do so, terms occurring in the top-k documents retrieved by the system in its initial run were selected as expansion candidates. In addition, they incorporated medical concepts that appeared in the discharge summary of each query. The main limitation of this approach was the utilization of resources that suffered from a restrained number of medical concepts. As such, many medical concepts could not be mapped to their corresponding terms in the given queries. To overcome shortcomings associated with the use of limited local resources, Zuccon et al. proposed using other external data sources [27]. In this context, the authors analyzed the results retrieved by two commercial web search engines (Google and Bing) on a set of queries formulated by laypeople to describe medical symptoms.

TABLE 1
SUMMARY OF REPRESENTATIVE RESEARCH WORKS

| Index | Approach | Strengths and Limitations |
|---|---|---|
| Zhu and Carterette. [33] | Medical Query Re-weighting | Strength: employing a variety of features for assigning weights for medical query concepts |
| | | Limitation: use of UMLS for mapping medical concepts which is not sufficient due to the limited domain coverage of this knowledge base |
| Martinez et al. [34] | UMLS-based Query Expansion | Strength: medical concept mapping performed with a trusted medical knowledge base |
| | | Limitation: semantic knowledge incompleteness in UMLS |
| Wang and Fang. [6] | Context-based Query Terms Expansion | Strength: locating all contexts in the original document collections that match the contexts of the key terms in verbose medical queries |
| | | Limitation: ignoring the semantics of key query terms and relying only on keyword overall to predict key term contexts |
| Stanton et al. [14] | Supervised Machine Learning based Query Reformulation | Strength: incorporating formal definitions of diseases using medical semantic resources |
| | | Limitation: ignoring other query term types that do not belong to symptom and disease categories, such as laboratory tests and medical devices |
| Shen et al.[24] | Bag-of-concepts based Medical Query Expansion | Strength: selecting concepts and their mapping entities in the used medical knowledge resources |
| | | Limitation: ignoring all non-medical query terms and limited domain coverage of used resources |
| Choi and Choi. [25] | Concept-based Medical Query Expansion | Strength: using discharge summary reports as a source for medical concept identification |
| | | Limitation: ignoring compound terms that may exist in the queries or medical documents and restricting the expansion process to discharge summary reports provided in the dataset |
| Goeuriot et al. [35] | Pseudo Relevance Feedback (PRF) based Query Reformulation | Strength: expanding query terms using relevance feedback obtained by initial system runs |
| | | Limitation: utilizing local medical resources with limited domain coverage |
| Zuccon et al. [27] | Web Search Engine based Query Expansion | Strength: exploiting web search engines to describe medical symptoms and expand medical query terms |
| | | Limitation: the used search engines are generic and do not perform as vertical domain-specific search systems |

The authors found that only three out of the top ten retrieved results by both search engines were marked as relevant. They concluded that existing commercial search engines cannot perform well when they are used in specific domains requiring expert knowledge such as the medical domain. We provide in Table 1 a summary of the representative research works discussed herein.

In order to address the above discussed limitations, we propose combining multiple medical semantic resources and query expansion techniques in a single MIR framework. Our attempt in this context is to bridge the semantic gap between medical queries and their corresponding medical documents. Accordingly, inspired by the strengths of previous approaches, we exploit trusted and well-recognized extrinsic medical resources, i.e. the UMLS lexicon and UMLS Metathesaurus, in our approach. Therefore, rather than using a limited resource such as the previously introduced discharge summary reports or a generic source such as web search engines, we use a combination of medical knowledge bases (such as MeSH, SNOMED, RxNorm [36]) for semantic concept highlighting and expansion. In addition, we propose a heuristical approach for re-weighting medical query terms based on their mappings to their relevant medical concepts in the used resources and experimentally demonstrate its impact.

### III. Detailed Description of the Proposed System

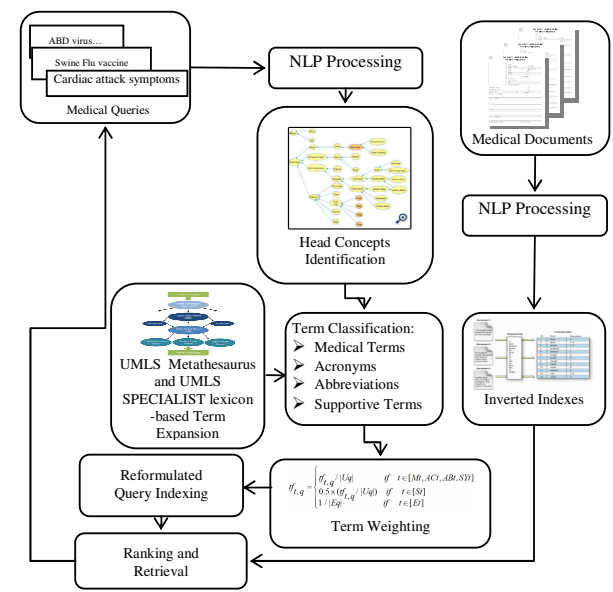

**FIGURE 1.** Block diagram of the proposed system architecture.

Figure 1 below depicts a block diagram for the main components of our proposed system, including their interactions. We detail in the remainder: a) the linguistic pre-processing step, b) the semantic processing of queries and their UMLS-based expansion, c) the semantic processing of documents and the construction of a semantic-based inverted index, d) the similarity computation between query and index documents and heuristic-based weighting strategy.

*A. Query Processing & UMLS-based Expansion*

Medical queries are verbose natural language queries that are inherently ambiguous and contain many terms that are hard to resolve. The fundamental goal is therefore to decipher their intent after highlighting and expanding their key components while removing less important entities that would impact retrieval performance. As depicted in Figure 1, a query is pre-processed through a sequence of NLP steps that include n-gram tokenization and stopword removal in order to filter out irrelevant terms [37]. We would like to point out that instead of only using a manually-constructed list of stopwords as proposed in [38], we utilize a term weighting scheme that employs the inverse document frequency [39] to assign term weights. Accordingly, in addition to the pre-defined list of stopwords, the proposed scheme assists in automatically constructing a list of additional stopwords that have weights below a threshold value $v$, which are obtained using Equation 1:

$$idf_t = \log_{10}\left(\frac{N}{df_t}\right) \quad (1)$$

where,
- $idf_t$ : is the inverse document frequency of a term $t$
- $N$ : is the total number of documents in the dataset
- $df_t$ : is an inverse measure of the informativeness of the term $t$

For the sake of generalization, the threshold value $v$ is automatically determined as follows. For each term $t$ that belongs to a document $d$, we obtain $idf\_list = \langle idf_{t1}, idf_{t2}, idf_{t3}, ..., idf_{tn} \rangle$, and find the maximum difference among the elements of the $idf\_list$ using $v = Max_{idf}(idf_{ti}, idf_{tj})$. Accordingly, all terms with weights less than $v$ are automatically added to the stopword list.

After this filtering step, a query $q$ is represented as:

$$q = (t_1, ..., t_n) \quad (2)$$

Where each $t_i$ belongs to any of the following categories [15]:
- Medical Terms, i.e. terms that can be mapped to medical concepts in the exploited medical semantic resources (e.g. the medical term *aortic*).
- Acronyms, e.g. *ARV* that stands for 'Adelaide River Virus', 'Average Rectified Value' or other medical terms.
- Abbreviations, e.g. *Abd* that stands for 'Abduction'.
- Supportive Terms, defined as any other terms in the query that could not be classified as acronyms, abbreviations, or medical terms (such as: *replacement, status…*).

To perform the category mapping step, we define a sliding window of length $n = 3$ for finding uni-, bi- and trigram tokens among query terms. This task is performed with the assistance of the exploited medical semantic resources rather than conventional methods based on statistical information such as term frequency and inverse document frequency [40], residual inverse document frequency and weighted information gain [41], and google n-gram term and query frequency [27]. In this context, we submit all n-gram tokens to the UMLS lexicon [20] in order to classify them into the four categories. In addition, we utilize the MetaMap tool to detect synonymous medical terms for any of the terms that fall under the three first categories based on the UMLS Metathesaurus; which includes data from MeSH, SNOMED, RxNorm, and other collections [36].

The multiple semantic resource based query processing scenario is formalized in Algorithm 1:

**Algorithm 1. User Query Processing Using UMLS Metathesaurus and UMLS Lexicon**

**Input**: User_Query
**Output**: List of medical terms in the user query,
List of medical acronyms and abbreviations in the user query,
List of synonyms for medical terms in the user query

01: Med_list ← ⟨ ⟩;
02: Temp_med_list ← ⟨ ⟩;
03: Syn_list ← ⟨ ⟩;
04: Acr_abbr_list ← ⟨ ⟩;
05: Med_list = **GET_MED_TERMS_USING_METAMAP**(User_Query);
06: Temp_med_list = **GET_MED_TERMS_USING_MRDEF**(User_Query);
07: **for** i←0; i < Temp_med_list.length; i++
08:     **ADD**(Med_list, Temp_med_list[i]);
09: **for** i←0; i < Med_list.length; i++
10:     **if HAS_SYNONYM**(Med_list[i]) then
11:         **ADD**(Syn_list, **GET_SYN_FROM_LEXICON**(Med_list[i]));
12: Acr_abbr_list = **GET_ACR_ABBR_FROM_LEXICON**(User_Query);

To demonstrate these steps, we consider the following two example medical queries that are obtained from two different datasets (CLEF e-Health 2014 and TREC).

- $q_1$: MRSA and wound infection, and its danger (QTRAIN2014.1 of CLEF e-health2014 dataset [35]).
- $q_2$: Patients diagnosed with localized prostate cancer and treated with robotic surgery (Number: 104 of TREC dataset topics 101-135.txt[2]).

TABLE 2
N-GRAM QUERY PROCESSING RESULTS

| Queries / N-grams | $q_1$ | $q_2$ |
|---|---|---|
| Ug | [mrsa, wound, infect, danger] | [patient, diagnos, prostat, cancer, robot, surgeri] |
| Bg | [mrsa wound, wound infect, infect danger] | [prostat cancer, robot surgery] |
| Tg | [mrsa wound infect, wound infect danger] | [local prostat cancer] |

After the stopword filtering step, query terms are stemmed using Porter stemmer [42]. We use the n-gram tokenization technique to highlight lists of uni-grams (*Ug*), bi-grams (*Bg*) and trigrams (*Tg*). The output for the example queries is shown in Table 2.

---
[2] trec.nist.gov/data/medical/11/topics101-135.txt

The UMLS lexicon is used to extract and expand medical acronyms and abbreviations in the user query through its ACRONYM table. We also utilize the UMLS lexicon to find the synonyms of all medical query terms by using the LEXSYNONYM table. The query is then enhanced by incorporating all of the full representations of the extracted acronyms and abbreviations, and also by including the extracted synonyms. The resulting lists are presented in Table 3.

TABLE 3
USING ACRONYM AND LEXSYNONYM TABLES

| Queries<br>Terms | Lists of terms for $q_1$ | Lists of terms for $q_2$ |
|---|---|---|
| Acronyms | [mrsa] | [] |
| Abbreviations | [] | [] |
| Expansion | [methicillin resistant staphylococcus aureus] | [prostate carcinoma, malignant neoplasm of prostate] |
| Synonyms | [vulnerat] | [suspect prostat cancer, robot assist surgeri] |

In addition, we employ the MetaMap tool which maps tokens to the UMLS Metathesaurus. It locates all UMLS concepts associated with terms in biomedical texts using the knowledge intensive method that is based on symbolic, NLP and computational linguistic techniques as detailed in [21]. The results of this step are the lists of medical terms $Mt_1$ and $Mt_2$ that are described below:

- For $q_1$: $Mt_1$ = [wound, infect]
- For $q_2$: $Mt_2$ = [patient, diagnos, robot surgery, local prostat cancer]

All remaining terms that are not recognized using the previous steps are considered as supportive terms. The supportive term lists $St_1$ and $St_2$ are as follows:

- For $q_1$: $St_1$ = [danger]
- For $q_2$: $St_2$ = [robot, treat]

Based on the previous steps, we update and expand the input queries resulting in the two queries $Eq_1$ and $Eq_2$:

- $Eq_1$: mrsa, wound, infect, danger, methicillin resistant staphylococcus aureus.
- $Eq_2$: patient, diagnos, prostat, cancer, robot, surgery, prostat cancer, robot surgery, local prostat cancer, suspect prostat cancer, robot assist surgery, prostate carcinoma, malignant neoplasm of prostate, treat

### B. Generation of Semantic-based Inverted Index Documents

In this context and unlike conventional approaches that use the bag-of-words model to index medical documents, we construct an inverted index that stores medical terms and their semantically-relevant terms that are obtained from the exploited medical semantic resources. A language resource pre-processing step consists in first applying the Jsoup[3] parser for cleaning and extracting textual content from the medical documents since they are provided as raw HTML web pages. We then resort to downcasing and removing stopwords as described in the previous section. Next, the Porter stemmer is utilized to stem each term in the remaining text. All stemmed terms are added to the inverted index. If a term represents a medical acronym or abbreviation, then we add all full representations of the term to the index document. Similarly, all compounds, i.e. bi- and trigram terms and their acronyms, are also added. Accordingly, the automatic construction of the inverted index documents is summarized in Algorithm 2.

```
Algorithm 2. Automatic Construction of Inverted Index Documents
Input:   Document collection
Output:  Semantically enhanced inverted Index

01:  xRaw_doc_list ← ⟨ ⟩ ;
02:  Processed_doc_list ← ⟨ ⟩ ;
03:  while (HAS_NEXT(Raw_doc_list))
04:     ADD (Processed_doc_list, JSOUP_PARSE(Raw_doc_list.next))
05:  while (HAS_NEXT(Processed_doc_list))
06:     Temp_doc= GET_NEXT(Processed_doc_list)
07:     CASE_FOLDING(Temp_doc)
08:     REMOVE_STOP_WORD(Temp_doc)
09:     PORTER_STEMMER(Temp_doc)
10:     for i←0; i < Temp_doc.length; i++
11:        ADD(Inverted_index, term)
12:        If HAS_ACR_ABBR_IN_LEXICON (term) then
13:           ADD(Inverted_index, EXTRACT_FULL_FORM(term))
14:     Bigrams_list = GET_BIGRAMS_FROM_DOC(Temp_doc)
15:     Trigrams_list = GET_TRIGRAMS_FROM_DOC(Temp_doc)
16:     While (HAS_NEXT(Bigrams_list))
17:        Temp_bigram = GET_NEXT(Bigrams_list)
18:        ADD(Inverted_index, Temp_bigram)
19:        If HAS_ACR_ABBR_IN_LEXICON (Temp_bigram) then
20:           ADD(Inverted_index, EXTRACT_FULL_FORM(Temp_bigram))
21:     While (HAS_NEXT(Trigrams_list))
22:        Temp_trigram = GET_NEXT(trigrams_list);
23:        ADD(Inverted_index, GET_NEXT(Temp_trigram));
24:        If HAS_ACR_ABBR_IN_LEXICON(Temp_trigram) then
25:           ADD(Inverted_index, EXTRACT_FULL_FORM(Temp_trigram))
26:  foreach(term:inverted_index)
27:     CALCULATE_TF(term)
28:     CALCULATE_TFIDF (term);
```

### C. Similarity Computation & Heuristic based Weighting

In order to determine the similarity between the expanded queries and the indexed medical documents, we use a state-of-the-art vector space model [43] to demonstrate the benefits of our semantic-based processing techniques. It employs the *tf–idf* weighting scheme to assign a weight for each term *t* in a document *d*. In our approach, we use the *Normalized–tf*$_{t,d}$ where term occurrences are usually normalized to prevent a bias towards longer documents (which may have a higher term count regardless of the actual importance of that term in the document) to give a

---
[3] https://jsoup.org/

measure of the importance of term *t* within a particular document *d*:

$$Normalized - tf_{t,d} = \begin{cases} tf_{t,d} / |d| & if \ tf_{t,d} > 0 \\ 0, & Otherwise \end{cases} \quad (3)$$

Where $tf_{t,d}$ is the number of occurrences of term *t* in *d*, and |*d*| is the length of document *d*.

We furthermore heuristically argue that medical terms, their synonyms, abbreviations and acronyms are more informative, essential for retrieval and have a higher degree of contribution to the meaning of the query. Therefore, we assign higher weights for these terms against other terms. Accordingly, expansion terms characterizing the full representations of medical acronyms and abbreviations get lower weights and supportive terms get the lowest weights among all query terms. We use the following heuristical formula for calculating the occurrences of query terms $tf_{t,q}$ to give a higher weight for medical terms, acronyms, abbreviations and medical synonyms, against other supportive terms and other semantically-related concepts added to the query:

$$tf_{t,q} = \begin{cases} \frac{tf_{t,q}}{|q|} & if \ t \text{ is a medical term, acronym, abbreviation or synonym} \\ \frac{tf_{t,q}}{2.|q|} & if \ t \text{ is a supportive term} \\ \frac{1}{|rq|} & if \ t \text{ is an other semantically-related concept} \end{cases} \quad (4)$$

Where |*q*| is the length of the original query and |*rq*| is the length of the reformulated query.

---

**Algorithm 3  Assignment of Final Weights to Query Terms based on their Category**

**Input**: rq_terms_list [t1, t2, ...,tn], q_terms_list [t1, t2, ..., tn], Acr_abbr_list [t1, t2, ...,tn], Syn_list [t1, t2, ...,tn], Med_list [t1,t2, ..., tn], Sup_list [t1,t2, ..., tn], Exp_list [t1, t2, ..., tn]

**Output**: Hashmap of expanded query terms with their weights.

```
01:   Weight_hmap ← ⟨ ⟩
02:   term_count = 0
03:   term_weight = 0
04:   for i←0; i < rq_terms_list.length; i++
05:      term_count = GET_TERM_COUNT(rq_terms_list[i])
06:      if Med_list.contains(rq_terms_list[i]) or
          Acr_abbr_list.contains(rq_terms_list[i]) or
          Syn_list.contains(rq_terms_list[i]) then
07:         term_weight = term_count / q_terms_list.length;
08:      else
09:         if Sup_list.contains(rq_terms_list[i]) then
10:            term_weight = 0.5 * (term_count / q_terms_list.length)
11:         else
12:            if Exp_list.contains(rq_terms_list[i]) then
13:               term_weight = 1 / rq_terms_list.length
14:      Weight_hmap.PUT(rq_terms_list[i],term_weight)
15:   return Weight_hmap
```

---

In Formula (4), we give medical synonyms in the expanded query the same weight as their semantically-related terms in the original query. For the example query $q_2$, the term 'localized prostate cancer' and its synonym 'malignant neoplasm of prostate' have the same weight. But, we reduce the weight of all other terms that are semantically related to the original query terms without being synonymous. As far as the example query $q_1$ is concerned, the term 'mrsa' is given a higher weight than its full representation 'methicillin resistant staphylococcus aureus'.

Algorithm 3 is used for assigning weights to query terms based on their category.

To illustrate this step, we apply the algorithm for queries $q_1$ and $q_2$ and its results are compiled in Table 4.

TABLE 4
ASSIGNMENT OF WEIGHTS FOR THE EXAMPLE QUERY TERMS

| Query | | Length | Terms in $Eq$ | Term Type | Weight |
|---|---|---|---|---|---|
| $Uq_1$ | mrsa, wound, infect, danger | 4 | Mrsa | Acronym | 0.25 |
| | | | Wound | Medical | 0.25 |
| | | | Infect | Medical | 0.25 |
| $Eq_1$ | mrsa, wound, infect, danger, methicillin resistant staphylococcus aureus, vulnerat | 6 | Danger | Supportive | 0.12 |
| | | | methicillin resistant staphylococcus aureus | Other | 0.16 |
| | | | Vulnerat | Synonym | 0.25 |
| | | | Mal | Synonym | 0.20 |
| $Uq_2$ | patient, diagnos, prostat, cancer, robot, surgery, prostat cancer, robot surgery, local prostat cancer | 9 | Patient | Medical | 0.11 |
| | | | Diagnos | Medical | 0.11 |
| | | | Prostat | Medical | 0.11 |
| | | | Cancer | Medical | 0.11 |
| $Eq_2$ | patient, diagnos, prostat, cancer, robot, surgery, prostat cancer, robot surgery, local prostat cancer, suspect prostat cancer, robot assist surgery, prostate carcinoma, malignant neoplasm of prostate, treat | 14 | Robot | Supportive | 0.05 |
| | | | Surgery | Medical | 0.11 |
| | | | prostat cancer | Medical | 0.11 |
| | | | robot surgery | Medical | 0.11 |
| | | | local prostat cancer | Medical | 0.11 |
| | | | suspect prostat cancer | Other | 0.07 |
| | | | robot assist surgery | Other | 0.07 |
| | | | prostate carcinoma | Synonym | 0.11 |
| | | | malignant neoplasm of prostate | Synonym | 0.11 |
| | | | Treat | Supportive | 0.05 |

After determining the *tf–idf* results for query and document terms, the cosine similarity model is used to find the semantic similarity between document $\vec{d}$ and reformulated query $\vec{rq}$ according to:

$$sim(d, rq) = \cos ine(d, rq) = \frac{\vec{d}.\vec{rq}}{|\vec{d}||\vec{rq}|} \quad (5)$$

The results of the scoring function are returned as a list of relevant medical documents that are ordered in a descending manner starting from the most relevant document (i.e. the first result with the highest number of matching terms).

## IV. Experimental Setup and Evaluation

### A. Experimental Setup

In order to carry out experiments and evaluate the quality of our proposed head concepts selection approach, we used the CLEF e-Health 2014 medical dataset that comprises verbose queries associated with their relevance judgments. In the same manner as proposed in [38], we decided to use this dataset rather than the TREC queries. The reason is because CLEF queries are less artificial than TREC queries and also are more informative than queries obtained from a web search query log, where users often provide short queries with a small number of keywords to express their information needs. The components of the dataset are:

- Medical documents that are acquired automatically from various medical web sites, including pages certified by the Health On the Net[4] and other well-known medical databases [35]. The dataset comprises around one million semi-structured medical documents in HTML format that are distributed over 8 .zip files; where each file contains multiple .dat files with different medical topics. Each file contains multiple documents with the following format as depicted in Figure 2:
    - #UID: unique identifier for each document
    - #DATE: date the document was obtained
    - #URL: URL for the source of the document
    - #CONTENT: the raw HTML content of web pages

**FIGURE 2.** Example of a medical document from the CLEF eHealth 2014 dataset.

- Verbose medical queries divided into one set of five training queries and one set of fifty test queries created by experts (i.e. registered nurses and clinical documentation researchers) involved in the CLEF e-Health consortium. Queries are created based on the main disorders diagnosed in a set of selected patients' discharge summaries. As depicted in Figure 3, queries have a standard format that includes the following elements:
    - id: a unique identifier for each query
    - discharge_summary: the resource queries have been built from
    - title: a short version of the user query
    - desc: the verbose form of the query in the title field
    - profile: a brief description about the patient who submitted the query
    - narr: expected content in relevant documents

---

[4] http://www.hon.ch/

**FIGURE 3.** Query example from CLEF eHealth 2014 dataset.

- Relevance judgments collected from professional assessors using Relevation[5]: a system designed to record relevance judgments for information retrieval evaluation [44]. It provides a web interface through which judges can upload their documents, queries and relevance assessments. Relevance grades are four-valued in the interval 0-3. The value 0 means that a document is irrelevant to a given query, while 1 refers to a document that is on topic of a given query but deemed unreliable. The values 2 and 3 refer to documents that are relevant to the given query where value 3 is assigned to the highest relevant documents. These relevance grades are mapped into a binary scale, with grades 0 and 1 corresponding to the binary grade 0 (irrelevant) and grades 2 and 3 corresponding to the binary grade 1 (relevant).

In order to experimentally validate our proposal and evaluate the quality of the produced results, we have performed several runs as described below:

1. Incremental runs starting from a baseline run in which we solely utilized a standard inverted index to a final run that incorporates the techniques covered in this article including the use of extrinsic semantic resources and heuristical term re-weighting. The idea of using the baseline run is to set an initial measure of process functionality before carrying out any modifications. The goal is to demonstrate the quantitative improvements brought about by enriching the search framework with the processing modules described in Section III.

2. Several comparative-based runs allowing us to compare the results produced by our system with those of three state-of-the-art systems using the same dataset.

As far as the incremental runs are concerned, we executed five different runs. First, we started with the baseline RUN 1 where we used only the primitive inverted index and basic NLP techniques for both query and document processing. Second, in RUN 2, we re-indexed the document collection by incorporating compound medical terms, their acronyms and abbreviations using both the UMLS lexicon and UMLS Metathesaurus. For query processing, we identifed both acronyms and abbreviations and included unigram tokens only. The full representations of both acronyms and abbreviations were then added to the expanded query. Next, we performed three additional runs developed based on RUN 2. In RUN 3, we incorporated compound terms in addition to their acronyms and abbreviations. In RUN 4, we used the UMLS Metathesaurus via the MetaMap tool for categorical classification (cf. Section III.A). In this context, we extracted medical terms and other supportive query terms, and assigned weights based on their categorization as discussed in Section III.C. Finally, in RUN 5, we used the UMLS lexicon to expand query terms by adding their synonyms that are recognized by the exploited semantic resources. Synonyms in this context were assigned higher weights against other semantically-relevant terms based on Formula (4). In Table 5 we provide a brief summary of each of these runs, in addition to the parameters used and their role during each run.

---

[5] http://ielab.github.io/relevation/

TABLE 5
EXPERIMENTAL METHODS USED IN EACH RUN

| RUN | Experimental Method | Main Features |
|---|---|---|
| 1 | Baseline (simple inverted index with cosine similarity retrieval model) | Bag-of-words model, used to build standard inverted indexes without incorporating any medical semantic resources or heuristical term re-weighting techniques |
| 2 | Semantically enhanced inverted index with acronyms and abbreviations obtained using the UMLS lexicon | Use of Acronyms and Abbreviations with the inclusion of unigram tokens only |
| 3 | Experimental methods used in RUN 2 in addition to incorporating compound terms | Utilization of Acronyms and Abbreviations in the form of uni-, bi- and trigram tokens |
| 4 | Experimental methods used in RUN 3 in addition to including medical terms obtained using UMLS Metathesaurus and query term reweighting techniques | Single and Compound Acronyms and Abbreviations, in addition to Medical terms obtained from the UMLS Metathesaurus |
| 5 | Experimental methods used in RUN 4 in addition to enriching the reformulated queries with medical synonyms obtained using the UMLS lexicon | All parameters used in RUN 4, in addition to synonymous concepts obtained from UMLS |

Similar to the evaluation of equivalent MIR systems, we considered using the Precision@10 (P@10) evaluation metric to assess the performance of our medical search framework. This metric is among the most commonly used metrics among web-scale information retrieval systems. The P@10 corresponds to the number of relevant results among the first page results (top 10 documents) that are retrieved by the system. Formally, the precision metric is defined as follows:

$$P = \frac{|\{relevant\ documents\} \cap \{retrieved\ documents\}|}{|\{retrieved\ documents\}|} \quad (6)$$

### B. Experimental Results

As depicted in Figure 4, for RUN 2, among the 55 test and training queries, one query (qtest2014.27) showed low quality results compared to the baseline run, 3 queries (qtest2014.3, qtest2014.433 and qtest2014.47) yielded better results and the rest 51 queries produced equal results to those in the baseline run. The main reason behind the low quality results of qtest2014.27 is that this query contains an acronym that has multiple full representation forms which lead to retrieving documents that are irrelevant to the query context.

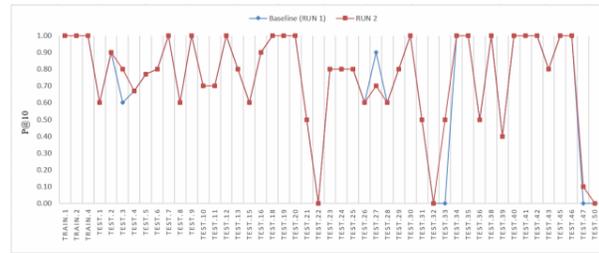

**FIGURE 4.** Baseline run against RUN 2 at P@10.

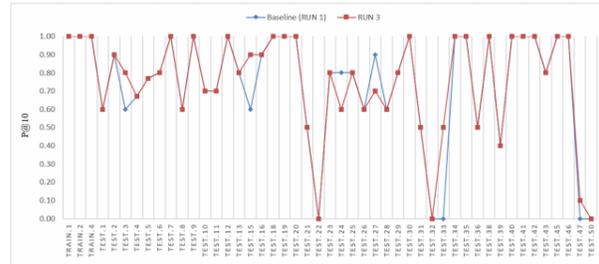

**FIGURE 5.** Baseline run against RUN 3 at P@10.

In Figure 5, we show the comparison between the results obtained from RUN 3 and those obtained from the baseline run. Here, we find 2 queries (qtest2014.24 and qtest2014.27) obtaining lower precision results than those in the baseline run, 4 queries (qtest2014.3, qtest2014.15, qtest2014.433 and qtest2014.47) yielding higher precision results and the other 49 queries with equal precision. The main reason for the low precision of the qtest2014.24 and qtest2014.27 queries lies in the fact that these queries contain bi- and trigrams that can be related to many contexts other than the query context only. They therefore returned a significant number of irrelevant documents, thereby decreasing the precision of the system.

The most significant improvement was achieved in RUN 4. As we can see in Figure 6, 11 queries among the 55 queries produced more precise results compared to their corresponding queries in the baseline run. This is mainly due to the utilization of the proposed heuristical re-weighting technique discussed in section III.C. In this context, query terms were classified and assigned different weights based on the exploited medical resources (i.e. the UMLS Metathesaurus via the MetaMap tool and the MRDEF table). Based on this run, we find that it is important to distinguish between query terms and assign higher weights to those that belong to the Acronym, Abbreviation, or Medical term categories. The results demonstrate that identifying terms that belong to these categories and enriching them with additional semantically-relevant medical terms produce more precise retrieval results. This leads to reducing the semantic gap between user queries and their corresponding medical documents in the dataset.

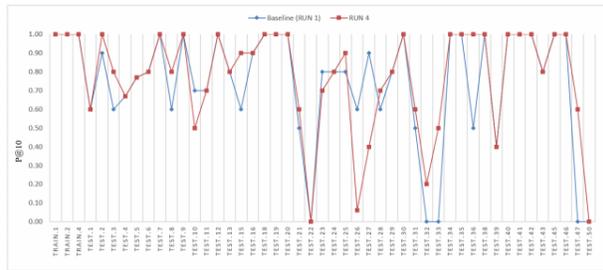

**FIGURE 6.** Baseline run against RUN 4 at P@10.

In the last run (RUN 5), we expanded the queries through adding medical synonyms that are related to the medical terms. Figure 7 shows a comparison between the P@10 results of RUN 5 and their counterparts of the baseline run. As we can see in this figure, further improvements in terms of precision were achieved, which confirms the findings in [3].

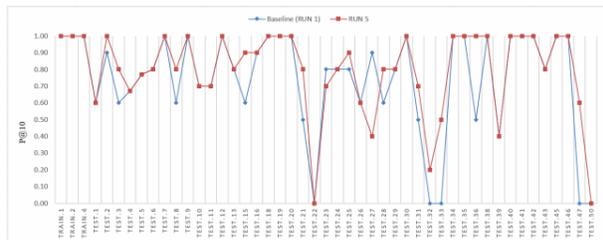

**FIGURE 7.** Baseline run against RUN 5 at P@10.

In Figure 8, we provide a comprehensive comparison between all system runs with the baseline run. These results represent the overall system effectiveness after testing all training and test queries provided in the CLEF e-health 2014 dataset.

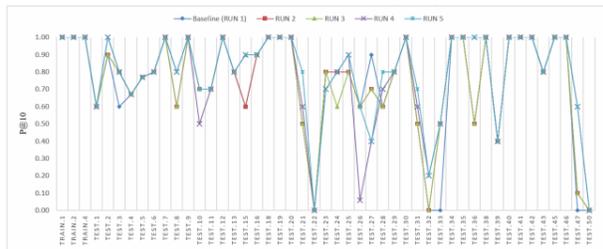

**FIGURE 8.** Comparison between all RUNs and the base-line run.

Next, we compare the results produced by our system to those produced by three CLEF participant teams who used the CLEF e-Health 2014 dataset for evaluating their proposed systems. These systems are:

- GRIUM [24] in their best run (EN-Run.5).
- SNUMEDINFO [25] in their best run (EN-Run.2).
- KISTI [45] in their best run (EN-Run.2).

**TABLE 6**
**COMPARISON WITH OTHER MIR SYSTEMS**

| P@10 | Our System | GRIUM | SNUMEDINFO | KISTI |
|---|---|---|---|---|
| Baseline Run | 0.7211 | 0.7180 | **0.7380** | 0.7300 |
| Best Run | **0.7945** | 0.7560 | 0.7540 | 0.7400 |

As shown in Table 6, the precision of the results produced by our proposed system is higher than those produced by the three compared systems (GRIUM, SNUMEDINFO and KISTI). The main reason for this improvement is due to the exploitation of n-grams, medical acronyms and abbreviations (using the employed medical semantic resources) in the indexing process. The three systems discussed in this article use basic indexing and retrieval algorithms provided in the Indri and Lucene frameworks. The authors of GRIUM proposed a retrieval model using a bag-of-concepts model rather than the traditional bag-of-words. They used the MetaMap tool for extracting medical concepts that exist in the user query to be considered in their query-document matching process. The main drawback of GRIUM lies in the fact that it ignores all query terms that are not identified as medical concepts by the MetaMap, which led to ignoring some important medical concepts such as medical acronyms and abbreviations. This explains their marginal improvement in terms of precision (i.e. 0.756) over the baseline run (i.e. 0.718). In SNUMEDINFO, the authors used a simple inverted index where compound terms were not included with the UMLS Metathesaurus for query expansion. The main drawbacks of SNUMEDINFO are: i) ignoring compound terms that may occur in both the document collection and user query, ii) ignoring the semantic nature of the extracted concepts (i.e. synonymy, meronymy...) from the UMLS Metathesaurus and, iii) giving all extracted concepts the same weight as in the original query. In addition, the authors did not tackle problems associated with acronyms and abbreviations in the documents and queries. Finally, the KISTI system demonstrated a slight improvement in terms of precision compared to the baseline technique. This is explained by the use of the related discharge summary information provided with each query in the dataset as their expansion resource. However, the discharge summary reports cannot be considered as a trusted medical semantic resource that can be used for expansion of medical queries because such reports were provided as example results only. It is important to point out that although our system was able to outperform the three systems in terms of precision, it still suffers from low computational efficiency running on an average configuration (i.e. PC with core I7 CPU-2.5 GHZ and 8GB of RAM) as it processes several large-scale semantic resources. We plan to address this issue in the upcoming prototype version through incorporating a classification module wherein medical documents as well as their corresponding queries will be classified under their relevant medical topics. In this context, instead of matching each query with every document in the dataset, we aim to find matches between queries and medical documents that fall under the same medical topic/s.

## V. Conclusions and Future Work

In this paper, we discussed the crucial role of medical semantic resources in addressing a key challenge for medical information retrieval systems; that is highlighting key medical concepts in queries and expanding them in order to decipher the query intent. Our aim in this context was to improve the query processing, document indexing and query-document matching step. We attempted to extend conventional methods that proposed to process queries using individual medical resources by employing a combination of extrinsic semantic resources, i.e. the MetaMap tool, MRDEF relational table and UMLS SPECIALIST lexicon. Additionally, we constructed semantically-enriched inverted files that captured key concepts, their acronyms, abbreviations as well as other semantically-relevant medical terms. As such, instead of only relying on representative key terms, our method constructs indexes that capture additional semantic dimensions which are utilized for matching, ranking and retrieval purposes. We also proposed to categorize query terms and assign higher weights to medical terms, their acronyms and abbreviations against other supportive terms. To validate our proposal, we evaluated the effectiveness of the proposed methods through developing a full-fledged prototype system comprising both query processing and document indexing techniques. We carried out several incremental runs on the one hand and comparative runs on the other hand against three state-of-the-art medical retrieval systems in order to quantify the impact of improvement achieved by the proposed system. Future developments include the incorporation of a query and document classification module in order to alleviate the current computational load. In addition, we noticed an important factor requiring further exploration: that is the consideration of semantic relations between the identified key medical concepts in the user queries. We believe that incorporating these relations will lead to better deciphering and understanding of the query intent and accordingly to the retrieval of more relevant results. This will definitely require updating/changing the current structure of the constructed inverted indexes into semantic networks that not only consist of head concepts, but also semantic and taxonomic relations that link them. In this context, we will update our query-document matching process to become a semantic-network based matching algorithm, wherein the closer the similarity between the networks, the higher the rank of the index document in the retrieval results.

## ACKNOWLEDGMENT
The authors would like to thank ELRA for offering the CLEFeHealth 2014 Task 3 Evaluation Package "CLEFeHealth 2014 Task 3 Evaluation Package, ELRA catalogue (http://catalog.elra.info), ISLRN: 725-020-897-275-7, ELRA ID: ELRA-E0043."

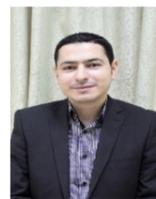

**Mohammed Maree**
received the Ph.D. degree in information technology from Monash University. He has published articles in various high-impact journals and conferences, such as ICTAI, Knowledge-Based Systems, and the Journal of Information Science. He is currently a Committee Member/Reviewer of several conferences and journals. He has supervised a number of master's students in the field of knowledge engineering, data analysis, information retrieval, natural language processing, and hybrid intelligent systems. He began his career as a Research and Development Manager with gSoft Technology Solution Inc. Then, he worked as the Director of Research and QA with Dimensions Consulting Company. Subsequently, he joined the Faculty of Engineering and Information Technology (EIT), Arab American University, Palestine (AAUP), as a full-time Lecturer. From September 2014 to August 2016, he was the Head of the Multimedia Technology Department, and from September 2016 to August 2018, he was the Head of the Information Technology Department. In addition to his work at AAUP, he worked as a Consultant for SocialDice and Dimensions Consulting companies. He is also the Head of the Multimedia Technology Department, Faculty of Engineering and Information Technology, AAUP.

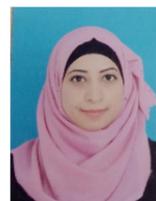

**Isra Noor**
received the B.S. degree (with honors) in computer information systems from An-Najah National University, Nablus, Palestine, in 2012 and the M.Sc. degree in computer science from the Arab American University, Jenin, Palestine, in 2017. She is currently the head of the Programming and Electronic Design Unit at An-Najah National University Hospital, Nablus, Palestine.

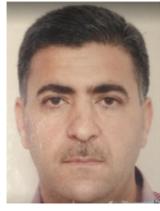

**Khalid S. Rabayah**
is an Associate Professor in Computer Science and Information Systems. He is the founder and the manger of an Information System research centre at the Arab American University-Jenin, Palestine. Currently Rabayah is an active researcher in three areas related to information systems and computing; Information systems, Wireless Networks and Internet of Things (IoT), and Data Analysis and Mining. His research interest in Information systems includes knowledge management, e-commerce, e-learning, technology adoption modelling and diffusion, especially in the context of developing countries, and cross-cultural issues in the use of IT. He authored and co-authored over 15 papers in this domain. His research interest in Wireless Networks and Internet of Things, started in 2014. Currently he is involved in a project which focuses on enhancing the quality of services (QoS) of Internet of things (IoT), based on light weight security protocols. Up till this date, he authored and co-authored 2 papers in this domain. His research interest in Data mining and data analysis started back in 2011. His research experience in data mining and analysis focuses on the use of the statistical software package of IBM SPSS and AMOS. He professionally uses these packages in developing predictive models for business processes and discovering hidden patterns in unstructured data and big data.

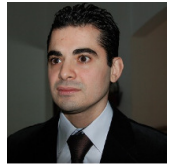

**Mohammed Belkhatir**
graduated from the University of Grenoble, France with an M.Phil and a Ph.D in Computer Science, both of which were supported by research grants from the French Ministry of Research. He is now an Associate Professor at the University of Lyon, France.

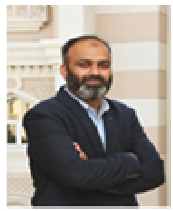

**Saadat M. Alhashmi**
received his PhD from Sheffield Hallam University, Sheffield, UK. Over the years, he has supervised a number of PhD students and published extensively in various high impact journals and conferences. Saadat joined the University of Sharjah, Sharjah, UAE in 2015 as an Associate Professor of MIS. His current research interests are business analytics, big data and the impact of technology on businesses.